\documentclass[12pt]{article}
\usepackage{amsmath}
\usepackage{graphicx}
\usepackage{bm}
\usepackage{fancyhdr}
\usepackage{amssymb}
\renewcommand{\theequation}{\arabic{section}.\arabic{equation}}

\begin{document}



\def\a{\alpha}
\def\b{\beta}
\def\d{\delta}
\def\e{\epsilon}
\def\g{\gamma}
\def\h{\mathfrak{h}}
\def\k{\kappa}
\def\l{\lambda}
\def\o{\omega}
\def\p{\wp}
\def\r{\rho}
\def\t{\tau}
\def\s{\sigma}
\def\z{\zeta}
\def\x{\xi}
\def\V={{{\bf\rm{V}}}}
 \def\A{{\cal{A}}}
 \def\B{{\cal{B}}}
 \def\C{{\cal{C}}}
 \def\D{{\cal{D}}}
\def\K{{\cal{K}}}
\def\O{\Omega}
\def\R{\bar{R}}
\def\T{{\cal{T}}}
\def\L{\Lambda}
\def\f{E_{\tau,\eta}(sl_2)}
\def\E{E_{\tau,\eta}(sl_n)}
\def\Zb{\mathbb{Z}}
\def\Cb{\mathbb{C}}

\def\R{\overline{R}}

\def\beq{\begin{equation}}
\def\eeq{\end{equation}}
\def\bea{\begin{eqnarray}}
\def\eea{\end{eqnarray}}
\def\ba{\begin{array}}
\def\ea{\end{array}}
\def\no{\nonumber}
\def\le{\langle}
\def\re{\rangle}
\def\lt{\left}
\def\rt{\right}

\newtheorem{Theorem}{Theorem}
\newtheorem{Definition}{Definition}
\newtheorem{Proposition}{Proposition}
\newtheorem{Lemma}{Lemma}
\newtheorem{Corollary}{Corollary}
\newcommand{\proof}[1]{{\bf Proof. }
        #1\begin{flushright}$\Box$\end{flushright}}

\baselineskip=20pt

\newfont{\elevenmib}{cmmib10 scaled\magstep1}
\newcommand{\preprint}{
   \begin{flushleft}
   \end{flushleft}\vspace{-1.3cm}
   \begin{flushright}\normalsize
   \end{flushright}}
\newcommand{\Title}[1]{{\baselineskip=26pt
   \begin{center} \Large \bf #1 \\ \ \\ \end{center}}}
\newcommand{\Author}{\begin{center}
   \large \bf
Kun Hao${}^{a}$,~Junpeng Cao${}^{b,c}$,~Guang-Liang Li${}^{d}$,~Wen-Li Yang${}^{a,e}\footnote{Corresponding author:
wlyang@nwu.edu.cn}$,
 ~ Kangjie Shi${}^a$ and~Yupeng Wang${}^{b,c}\footnote{Corresponding author: yupeng@iphy.ac.cn}$
 \end{center}}
\newcommand{\Address}{\begin{center}

     ${}^a$Institute of Modern Physics, Northwest University,
     Xian 710069, China\\
     ${}^b$Beijing National Laboratory for Condensed Matter
           Physics, Institute of Physics, Chinese Academy of Sciences, Beijing
           100190, China\\
     ${}^c$Collaborative Innovation Center of Quantum Matter, Beijing,
     China\\
     ${}^d$Department of Applied Physics, Xian Jiaotong University, Xian 710049, China\\
     ${}^e$Beijing Center for Mathematics and Information Interdisciplinary Sciences, Beijing, 100048,  China

   \end{center}}

\preprint \thispagestyle{empty}
\bigskip\bigskip\bigskip

\Title{A representation basis  for the  quantum integrable spin chain associated with the $su(3)$ algebra} \Author

\Address \vspace{1cm}

\begin{abstract}
An orthogonal basis of the Hilbert space for the  quantum spin chain associated with the $su(3)$ algebra is introduced. Such kind of basis could be treated as a nested
generalization of separation of variables (SoV) basis for high-rank quantum integrable models. It is found that  all the monodromy-matrix elements acting on a basis vector take simple forms. With the help of the basis, we construct  eigenstates of the $su(3)$ inhomogeneous spin torus
(the trigonometric $su(3)$ spin chain with antiperiodic  boundary condition) from its spectrum obtained via the off-diagonal Bethe Ansatz (ODBA). Based on small sites  (i.e. $N=2$) check, it is conjectured that the homogeneous limit of the eigenstates exists, which gives rise to the corresponding eigenstates of the homogenous model.

\vspace{1truecm} \noindent {\it PACS:} 75.10.Pq, 03.65.Vf, 71.10.Pm


\noindent {\it Keywords}: Spin chain; Bethe
Ansatz; $T-Q$ relation
\end{abstract}

\newpage



\section{Introduction}
\label{intro} \setcounter{equation}{0}

Quantum integrable system  has played an important role in understanding the physical
contents of the planar ${\cal{N}}=4$ super-symmetric Yang-Mills (SYM) theory and the planar
AdS/CFT \cite{Mal98, Bei12} (see also references therein). Moreover, it has also provided valuable insight into important
universality class in condensed matter
physics \cite{Duk04} and  cold atom systems \cite{Gua13}. In the past several decades, the integrable quantum spin chains
with $U(1)$-symmetry (with periodic boundary or with diagonal open boundaries \cite{Skl88}) and with some constrained
open boundaries \cite{Fan96,Nep04,Cao03,Gie05,Yan04-1,Gie05-1,Doi06,Baj06,Yan06-1} have been extensively studied by various Bethe ansatz methods
for a finite lattice and by the vertex operator method \cite{Jim94} in an infinite or a half-infinite lattice \cite{Jim92, Dav93, Jim95, Bas14}.

Very recently, an important progress has been achieved in solving the eigenvalue problem of integrable models without $U(1)$-symmetry \cite{Cao1} (i.e.,
the off-diagonal Bethe Ansatz (ODBA), for comprehensive introduction we refer the reader to \cite{Wan15}).
Several long-standing models \cite{Cao1,Cao2,Li14,Cao14,Hao14} have since been solved. It should be noted  that besides ODBA \cite{Cao-14-Bethe-state}
some other methods such as the q-Onsager algebra method \cite{Bas1,Bas3}, the separation of variables (SoV) method
\cite{1Fra08,Nic12,Nic13,Nic14} and the modified algebraic Bethe  ansatz method \cite{Bel13,Bel15,Bel15-1,Ava15} were also used to obtain the eigenstates of the XXZ spin
chains with generic boundary conditions. Remarkably, ODBA allows us to obtain  eigenvalues of the $U(1)$-broken models associated
with higher-rank algebras such as the $su(n)$ spin chain with generic integrable boundary fields \cite{Cao14}, the Izergin-Korepin model\footnote{It is a  model beyond A type.}  with generic boundary conditions \cite{Hao14},
the Hubbard model\cite{Li14} and the supersymmetric $t-J$ model \cite{Zha14} with unparallel boundary fields, and the open chain related to AdS/CFT \cite{Zha15}. However, the corresponding eigenstates for these models are still missing.

According to Liouville's theorem, a key feature of integrable models is that their variables are completely separable. This concept was generalized to quantum integrable models by Sklyanin \cite{Skl92} and provided a promising approach to construct eigenstates of quantum integrable models without $U(1)$-symmetry. Nevertheless, Sklyanin's SoV procedure has only succeeded for some rank-one quantum integrable models and a proper SoV scheme for the high-rank quantum integrable models is still absent. The main task of the present paper is to propose a nested SoV basis for the $su(n)$ spin chain model. As an example of application, we construct exact eigenstates of the $su(3)$ spin torus (i.e., the trigonometric $su(3)$ spin chain with anti-periodic boundary condition), an archetype high-rank quantum integrable model without highest weight reference state,  based on its spectrum recently obtained in \cite{Hao16}  via ODBA.

The paper is organized as follows. Section 2 serves as an introduction to our notations for
the inhomogeneous $su(n)$ spin torus and its spectrum. In section 3, we introduce a nested SoV basis of the Hilbert space of the $su(3)$ spin chain.
It is found that the actions of the monodromy matrix elements on a basis vector
have no compensating exchange terms on the level of the local operators (i.e.,
polarization free) and therefore  become drastically simple. In section 4, with the help of the basis,
as an example, we construct eigenstates of the transfer matrix for the $su(3)$ spin torus from its spectrum obtained via ODBA \cite{Hao16}. Concluding remarks are given in section 5.
Some detailed technical proofs are given in Appendices $A-D$.

\section{$su(n)$ spin torus and its spectrum}
\setcounter{equation}{0}

Let ${\rm\bf V}$ denote an $n$-dimensional linear space with an
orthonormal basis $\{|i\rangle|i=1,\cdots,n\}$. We introduce the
Hamiltonian $H$ as follows: \bea H=\sum_{j=1}^Nh_{j,j+1},\label{Ham}
\eea where $N$ is the number of sites and $h_{j,j+1}$ is the local
Hamiltonian given by \bea h_{j,j+1}=\frac{\partial}{\partial
u}\lt.\lt\{P_{j,j+1}\,R_{j,j+1}(u)\rt\}\rt|_{u=0}. \eea Here
$P_{j,j+1}$ is the permutation operator on the tensor space  and the
$R$-matrix $R(u)\in {\rm End}({\rm\bf V}\otimes {\rm\bf V})$ is the
trigonometric $R$-matrix  associated with the quantum group
\cite{Cha94} $U_q(\widehat{su(n)})$, which was first proposed by Perk and
Shultz \cite{Perk81} and further studied in
\cite{Perk83,Schu83,Perk06,Baz85,Jim86}\footnote{The $R$-matrix
given by (\ref{R-matrix-1}) corresponds to  the so-called principal
gradation, which is related to the $R$-matrix in homogeneous
gradation by some gauge transformation \cite{Nep02}.}  \bea R(u) &=&
\sinh({u} + \eta) \sum_{k=1}^{n} E^{k\,, k}\otimes E^{k\,, k}+
\sinh{u}
\sum_{k \ne l}^n E^{k\,, k}\otimes E^{l\,, l} \no \\
&&+ \sinh\eta \left(\sum_{k<l}^ne^{\frac{n-2(l-k)}{n}u} +
\sum_{k>l}^n e^{-\frac{n-2(k-l)}{n}u} \right) E^{k\,, l}\otimes
E^{l\,, k} ,\label{R-matrix-1} \eea where the $n^2$ fundamental
matrices $\{E^{k,l}|k,l=1,\cdots,n\}$ are all $n\times n$ matrices
with matrix entries
$(E^{k,l})^{\a}_{\b}=\delta^k_{\a}\,\delta^l_{\b}$ and $\eta$ is the
crossing parameter. The $R$-matrix satisfies the quantum Yang-Baxter
equation (QYBE)
\begin{eqnarray}
 R_{12}(u_1-u_2)R_{13}(u_1-u_3)R_{23}(u_2-u_3)=
 R_{23}(u_2-u_3)R_{13}(u_1-u_3)R_{12}(u_1-u_2), \label{QYB}
\end{eqnarray}
and possesses the properties:
\begin{eqnarray}
 &&\hspace{-1.45cm}\mbox{
 Initial condition}:\hspace{42.5mm}R_{12}(0)= \sinh\eta P_{1,2},\label{Initial}\\
 &&\hspace{-1.5cm}\mbox{
 Unitarity}:\hspace{11.5mm}R_{12}(u)R_{21}(-u)= \rho_1(u)\times{\rm id},\quad \rho_1(u)=-\sinh(u+\eta)\sinh(u-\eta),\label{Unitarity}\\
 &&\hspace{-1.5cm}\mbox{
 Crossing-unitarity}:\,
 R^{t_1}_{12}(u)R_{21}^{t_1}(-u-n\eta)
 =\rho_2(u)\times\mbox{id},\; \rho_2(u)=-\sinh u\sinh(u+n\eta),
 \label{crosing-unitarity}\\
 &&\hspace{-1.4cm}\mbox{Fusion conditions}:\hspace{22.5mm}\, R_{12}(-\eta)=-2\sinh\eta P^{(-)}_{1,2}.\label{Fusion}
\end{eqnarray}
Here $R_{21}(u)=P_{1,2}R_{12}(u)P_{1,2}$;
$P^{(-)}_{1,2}$  is the q-deformed anti-symmetric
 project operator \cite{Hao16} in the tensor product space  ${\rm\bf
V} \otimes {\rm\bf V} $; and $t_i$ denotes the transposition in the
$i$-th space. Here and below we adopt the standard notation: for any
matrix $A\in {\rm End}({\rm\bf V})$, $A_j$ is an embedding operator
in the tensor space ${\rm\bf V}\otimes {\rm\bf V}\otimes\cdots$,
which acts as $A$ on the $j$-th space and as an identity on the
other factor spaces; $R_{ij}(u)$ is an embedding operator of
$R$-matrix in the tensor space, which acts as an identity on the
factor spaces except for the $i$-th and $j$-th ones. For the $su(3)$ case the $R$-matrix reads
\bea\label{R-matrix-su3q}
R(u)=
\left(
  \begin{array}{ccc|ccc|ccc}
 \bar a(u) &    &    &    &    &    &    &    &   \\
      &\bar b(u)&    &\bar c(u)&    &    &    &    &   \\
      &    &\bar b(u)&    &    &    &\bar d(u)&    &   \\
\hline
      &\bar d(u)&    &\bar b(u)&    &    &    &    &   \\
      &    &    &    &\bar a(u)&    &    &    &   \\
      &    &    &    &    &\bar b(u)&    &\bar c(u)&   \\
\hline
      &    &\bar c(u)&    &    &    &\bar b(u)&    &    \\
      &    &    &    &    &\bar d(u)&    &\bar b(u)&    \\
      &    &    &    &    &    &    &    &\bar a(u)\\
\end{array}
\right), \eea where the matrix elements are \bea
&\displaystyle \bar a(u)=\sinh(u+\eta),\quad &\bar b(u)=\sinh u,\no\\
&\displaystyle \bar c(u)=e^{{u\over3}}\sinh\eta,\quad &\bar d(u)=e^{-{u\over3}}\sinh\eta.
\eea

Let us introduce the $n\times n$ twist matrix $g$
\bea
g=\lt(\begin{array}{cccc}&&&1\\
1&&&\\
&\ddots&&\\
&&1&\\
\end{array}\right),\quad {\rm and}\,\,g^n=1.\label{g-matrix}
\eea
For the $su(3)$ case, it reads
\bea
g=\lt(\begin{array}{ccc}0&0&1\\
1&0&0\\
0&1&0
\end{array}\right),\quad {\rm and}\,\,g^3=1.\label{g-matrix-3}
\eea
It is found that the $R$-matrix (\ref{R-matrix-1}) is invariant with $g$,
\bea
g_0\,g_{0^\prime}R_{00^\prime}(u)\,g^{-1}_0\,g^{-1}_{0^\prime}=R_{00^\prime}(u).\label{Invariant-R}
\eea
This property enables us to construct the integrable $su(n)$ spin torus model \cite{Hao16}.

Similar to the $su(2)$ spin torus (or the XXZ spin chain with anti-periodic boundary condition) \cite{Bat95},
the $su(n)$ spin torus is described by the Hamiltonian $H$ given by (\ref{Ham}) with  anti-periodic
boundary conditions
\bea
E^{k,l}_{N+1}=g_1\,E^{k,l}_1\,g_1^{-1},\quad k,l=1,\cdots,n.\label{anti-boundary}
\eea
Let us introduce the ``row-to-row" monodromy matrix
$T(u)$, an $n\times n$ matrix with operator-valued elements
acting on ${\rm\bf V}^{\otimes N}$, \bea T_0(u)
=R_{0N}(u-\theta_N)R_{0\,N-1}(u-\theta_{N-1})\cdots
R_{01}(u-\theta_1).\label{Monodromy-1} \eea Here
$\{\theta_j|j=1,\cdots,N\}$ are generic free complex parameters
usually called as inhomogeneity parameters.
The transfer matrix $t(u)$ of the associated spin chain with antiperiodic
boundary condition (\ref{anti-boundary}) can be constructed similarly as  \cite{Bat95}
\begin{eqnarray}
t(u)&=&tr_0\lt\{g_0\,T_0(u)\rt\}.\label{transfer}
\end{eqnarray}
The QYBE and the relation (\ref{Invariant-R})
lead to the fact that the transfer matrices $t(u)$ given by (\ref{transfer}) with
different spectral parameters are mutually commuting:
$[t(u),t(v)]=0$. The  Hamiltonian (\ref{Ham}) with the
anti-periodic boundary condition (\ref{anti-boundary})
can be obtained from the transfer matrix as
\begin{eqnarray}
&&H=\sinh\eta\, \frac{\partial \ln t(u)}{\partial u}|_{u=0,\{\theta_j\}=0}.
\end{eqnarray}

The eigenvalues $\L(u)$ of the transfer matrix $t(u)$ in case of $su(3)$ are given in terms of an inhomogeneous $T-Q$ relation \cite{Hao16}
\bea
&&\hspace{-1.2truecm}\L(u)=e^{\frac{u}{3}}\lt\{e^{\phi_1}e^u a(u)\frac{Q^{(1)}(u-\eta)}{Q^{(2)}(u)}
       +e^{-\phi_1}\omega e^{-u-{2\eta\over3}} d(u)\frac{Q^{(2)}(u+\eta)Q^{(3)}(u-\eta)}{Q^{(1)}(u)Q^{(4)}(u)}\rt.\no\\[4pt]
&&\hspace{-1.2truecm}\quad\quad\quad+\omega^2 e^{-u-{4\eta\over3}} d(u)\frac{Q^{(4)}(u+\eta)}{Q^{(3)}(u)}
       +a(u)d(u)\frac{Q^{(3)}(u-\eta)f_1(u)}{Q^{(1)}(u)Q^{(2)}(u)}\no\\[4pt]
&&\hspace{-1.2truecm}\quad\quad\quad
       +a(u)d(u)\lt.\frac{Q^{(2)}(u+\eta)f_2(u)}{Q^{(3)}(u)Q^{(4)}(u)} \rt\},\label{T-Q-1}
\eea
where
\bea
&&a(u)=\prod_{l=1}^N\sinh(u-\theta_l+\eta),\quad d(u)= \prod_{l=1}^N\sinh(u-\theta_l)=a(u-\eta),\label{ad-functions}\\
&&Q^{(i)}(u)=\prod_{l=1}^{N}\sinh(u-\l^{(i)}_l),\quad i=1,2,3,4,\no
\eea
$\omega=e^{\frac{2i\pi}{3}}$ and the functions $f_1(u)$ and $f_2(u)$ are given by
\bea
f_1(u)=f_1^{(+)}e^u+f_1^{(-)}e^{-u},\quad f_2(u)=f_2^{(-)}e^{-u}.\no
\eea
The $4N+4$ parameters $\{\l^{(i)}_l|l=1,\cdots, N;\,i=1,2,3,4\}$, $f_1^{(\pm)}$, $f_2^{(-)}$ and $e^{\phi_1}$ satisfy the associated
BAEs:
\bea
&&\omega e^{-\phi_1}e^{-\lambda^{(1)}_j-{2\eta\over3}} \frac{Q^{(2)}(\lambda^{(1)}_j+\eta)}{Q^{(4)}(\lambda^{(1)}_j)}
+a(\lambda^{(1)}_j)\frac{f_1(\lambda^{(1)}_j)}{Q^{(2)}(\lambda^{(1)}_j)}=0,\quad j=1,\cdots,N,\label{BAE-1}\\[4pt]
&&e^{\phi_1}e^{\lambda^{(2)}_j}Q^{(1)}(\lambda^{(2)}_j-\eta)
+d(\lambda^{(2)}_j)
\frac{Q^{(3)}(\lambda^{(2)}_j-\eta)f_1(\lambda^{(2)}_j)}{Q^{(1)}(\lambda^{(2)}_j)}=0,
\quad j=1,\cdots,N,\label{BAE-2}\\[4pt]
&&\omega^2 e^{-\lambda^{(3)}_j-{4\eta\over3}}Q^{(4)}(\lambda^{(3)}_j+\eta)
+a(\lambda^{(3)}_j)
\frac{Q^{(2)}(\lambda^{(3)}_j+\eta)f_2(\lambda^{(3)}_j)}{Q^{(4)}(\lambda^{(3)}_j)}=0,
\quad j=1,\cdots,N,\label{BAE-3}\\[4pt]
&&\omega
e^{-\phi_1}e^{-\lambda^{(4)}_j-{2\eta\over3}}\frac{Q^{(3)}(\lambda^{(4)}_j-\eta)}{Q^{(1)}(\lambda^{(4)}_j)}
+a(\lambda^{(4)}_j)\frac{f_2(\lambda^{(4)}_j)}{Q^{(3)}(\lambda^{(4)}_j)}=0,
\quad j=1,\cdots,N,\label{BAE-4} \eea \bea
&&\hspace{-1.2truecm}e^{\phi_1}e^{-\Theta-\chi^{(1)}+\chi^{(2)}}+e^{-2\Theta+\chi^{(1)}+\chi^{(2)}-\chi^{(3)}}f^{(+)}_1=0,\label{BAE-5}\\[4pt]
&&\hspace{-1.2truecm}\omega e^{-\phi_1}e^{-{2\eta\over3}+\Theta-\chi^{(1)}+\chi^{(2)}+\chi^{(3)}-\chi^{(4)}}
+\omega^2 e^{-{4\eta\over3}+\Theta-\chi^{(3)}+\chi^{(4)}-N\eta}\no\\[4pt]
&&+e^{2\Theta-N\eta}\lt\{e^{-\chi^{(1)}-\chi^{(2)}+\chi^{(3)}+N\eta}f^{(-)}_1
+e^{+\chi^{(2)}-\chi^{(3)}-\chi^{(4)}-N\eta}f^{(-)}_2\rt\}=0,\label{BAE-6}\\[4pt]
&&\hspace{-1.2truecm}\omega e^{-\Theta-\chi^{(3)}+\chi^{(4)}}+\omega^2 e^{\phi_1}e^{-{2\eta\over3}
-\Theta-\chi^{(1)}+\chi^{(2)}+\chi^{(3)}-\chi^{(4)}+N\eta}\no\\[4pt]
&&+e^{-2\Theta+N\eta}\lt\{\omega^2 e^{-{2\eta\over3}+\chi^{(1)}+\chi^{(2)}-\chi^{(4)}}f^{(+)}_1
+e^{\phi_1}e^{{2\eta\over3}-\chi^{(1)}+\chi^{(3)}+\chi^{(4)}+N\eta}f^{(-)}_2\rt\}=0,\label{BAE-7}\\[4pt]
&&\hspace{-1.2truecm}e^{-\phi_1}e^{-{4\eta\over3}+\Theta-\chi^{(1)}+\chi^{(2)}-N\eta}+\omega^2
e^{-{2\eta\over3}+2\Theta-\chi^{(1)}-\chi^{(2)}+\chi^{(4)}-N\eta}f^{(-)}_1=0,\label{BAE-8}
\eea
where
\bea
\Theta=\sum_{l=1}^N\theta_l,\quad \chi^{(i)}=\sum_{l=1}^N\l^{(i)}_l,\quad i=1,2,3,4.
\eea

In homogeneous limit: $\{\theta_j\rightarrow 0\}$, the resulting $T-Q$ relation (\ref{T-Q-1}) and the associated
BAEs (\ref{BAE-1})-(\ref{BAE-8}) give rise to the eigenvalue and BAEs of the corresponding homogeneous spin chain
(i.e., the $su(3)$ spin torus).

\section{Nested SoV basis}
\setcounter{equation}{0}
In this section, we propose a convenient basis of the Hilbert space  parameterized by the $N$ generic inhomogeneity parameters
$\{\theta_j|j=1,\cdots,N\}$. It is found that actions of all the monodromy matrix elements on a basis vector
take drastically simple forms  like those in the so-called F-basis\footnote{It is interesting to study the relation between this basis and the
F-basis \cite{Alb00,Yan06}.} \cite{Dri83,Mai96,Alb00,Alb01,Yan06}. All these ingredients allow us to construct exact eigenstates of the $su(n)$ spin torus model.

For convenience, let us introduce the notations
\bea
&&A(u)=T^1_1(u),\quad B_i(u)=T^1_i(u),\quad C^i(u)=T^i_1(u), \quad {\rm for}\,\, i=2,\ldots,n,\label{A-operator}\\[4pt]
&&D^i_j(u)=T^i_j(u),\quad {\rm for}\,\, i,j=2,\ldots,n.\label{D-operator}
\eea
The  exchange relations among the above operators are listed in Appendix A.
Let us introduce further the left quasi-vacuum state $\langle 0|$ and the right quasi-vacuum state $ |0\rangle$
\bea
\langle 0|=\langle 1,\cdots,1|,\quad |0\rangle=|1,1,\cdots,1\rangle.\label{left-vacuum}
\eea The operators (\ref{A-operator})-(\ref{D-operator}) acting on the states give rise to
\bea
&&\langle 0|\,A(u)=a(u)\,\langle 0|, \quad\langle 0|\,D^l_i(u)=d(u)\,\d^l_i\,\langle 0|,\quad i,l=2,\cdots,n, \label{D-action}\\[4pt]
&&\langle 0|\,B_i(u)=0,\quad \langle 0|\,C^i(u)\neq 0,\quad i=2,\cdots,n,\label{B-action}\\[4pt]
&&A(u)\,|0\rangle=a(u)\,|0\rangle,\quad D^l_i(u)\,|0\rangle=d(u)\,\d^l_i\,|0\rangle,\quad i,l=2,\cdots,n, \label{A-action-1}\\[4pt]
&&C^i(u)\,|0\rangle=0,\quad B_i(u)\,|0\rangle\neq 0,\quad i=2,\cdots,n,\label{B-action-1}
\eea
where the functions $a(u)$ and $d(u)$ are given by (\ref{ad-functions}).

In the following part of this section, taking the $su(3)$ spin chain as an example, we construct a nested SoV basis of the Hilbert space. The generalization to the
$su(n)$ case is given in Appendix B. For two non-negative integers $m_2$ and $m$  such that $m_2\leq m\leq N$, let us introduce $m$ positive integers $P=\{p_1,\cdots,p_m\}$ such that
\bea
1\leq p_1<p_2<\cdots<p_{m_2}\leq N,\quad 1\leq p_{m_2+1}<\cdots<p_m\leq N,\quad  {\rm and}\quad p_j\neq p_l.\label{P-condition}
\eea For each $P$ satisfies the above condition, let us introduce  left and right states parameterized by the $N$
inhomogeneity parameters $\{\theta_j\}$ as follows:
\bea
&&\hspace{-1.2truecm}\langle \theta_{p_1},\cdots,\theta_{p_{m_2}};\theta_{p_{m_2+1}},\cdots,\theta_{p_m}| =
\langle 0|C^2(\theta_{p_{1}})\cdots C^2(\theta_{p_{m_2}})\,C^3(\theta_{p_{m_2+1}})\cdots
C^3(\theta_{p_{m}}), \label{left-state-su(3)}\\[4pt]
&&\hspace{-1.2truecm}|\theta_{p_1},\cdots,\theta_{p_{m_2}};\theta_{p_{m_2+1}},\cdots,\theta_{p_m}\rangle=
B_3(\theta_{p_{m}})\cdots B_3(\theta_{p_{m_2+1}}) \,B_2(\theta_{p_{m_2}})\cdots B_2(\theta_{p_{1}})|0\rangle,
\label{right-state-su(3)}
\eea
where $m_2$ (resp. $m-m_2$) is the number of the operators $C^2(u)$ or $B_2(u)$
(resp. $C^3(u)$ or $B_3(u)$).

It is  easy to check that the states
(\ref{left-state-su(3)}) and the states (\ref{right-state-su(3)}) are eigenstates of the operator $D^3_3(u)$, namely,
\bea
&&\hspace{-1.2truecm}\langle \theta_{p_1},\cdots,\theta_{p_{m_2}};\theta_{p_{m_2+1}},\cdots,\theta_{p_m}|
\,D^3_3(u)=d(u)\prod_{l=m_2+1}^{m}\frac{\sinh(u-\theta_{p_{l}}+\eta)}
{\sinh(u-\theta_{p_{l}})}\no\\[2pt]
&&\quad\quad\times \langle \theta_{p_1},\cdots,\theta_{p_{m_2}};\theta_{p_{m_2+1}},\cdots,\theta_{p_m}|,\label{Eigenvalue-left-3}\\[4pt]
&&\hspace{-1.2truecm}D^3_3(u)\,|\theta_{p_1},\cdots,\theta_{p_{m_2}};\theta_{p_{m_2+1}},\cdots,\theta_{p_m}\rangle
=d(u)\prod_{l=m_2+1}^{m}\frac{\sinh(u-\theta_{p_{l}}+\eta)}
{\sinh(u-\theta_{p_{l}})}\no\\[2pt]
&&\quad\quad\times | \theta_{p_1},\cdots,\theta_{p_{m_2}};\theta_{p_{m_2+1}},\cdots,\theta_{p_m}\rangle.\label{Eigenvalue-right-3}
\eea
Noting the fact that $d(\theta_l)=0,\,l=1,\cdots,N$ and  using the exchange relations (\ref{CD})-(\ref{TT-3}), we can derive some useful relations
\bea
&&\hspace{-1.2truecm}\langle \theta_{p_1},\cdots,\theta_{p_{m_2}};\theta_{p_{m_2+1}},
\cdots,\theta_{p_m}|\,D^i_j(\theta_{p_l})=0,\quad l=m+1,\cdots,N,\,{\rm and}\,\, i,j=2,3,\label{Vanshing-1}\\[4pt]
&&\hspace{-1.2truecm}\langle \theta_{p_1},\cdots,\theta_{p_{m_2}};\theta_{p_{m_2+1}},\cdots,\theta_{p_m}|\, B_i(\theta_{p_l})=0,
\quad l=m+1,\cdots,N,\,{\rm and}\,\, i=2,3, \label{Vanshing-2}\\[4pt]
&& \hspace{-1.2truecm}D^i_j(\theta_{p_l})|
\theta_{p_1},\cdots,\theta_{p_{m_2}};\theta_{p_{m_2+1}},\cdots,\theta_{p_m}\rangle=0,\quad
l=m+1,\cdots,N,\,{\rm and}\,\, i=2,3,
\label{Vanshing-3}\\[4pt]
&&\hspace{-1.2truecm}C^i(\theta_{p_l})|
\theta_{p_1},\cdots,\theta_{p_{m_2}};\theta_{p_{m_2+1}},\cdots,\theta_{p_m}\rangle=0,\quad
l=m+1,\cdots,N,\,{\rm and}\,\, i=2,3. \label{Vanshing-4} \eea The
above relations and the exchange relations (\ref{CD})-(\ref{TT-3})
allow us to derive the  orthogonal relations between the left states
and the right states \bea &&\langle
\theta_{p_1},\cdots,\theta_{p_{m_2}};\theta_{p_{m_2+1}},\cdots,\theta_{p_m}|
\theta_{q_1},\cdots,\theta_{q_{m'_2}};\theta_{q_{m'_2+1}},\cdots,\theta_{q_{m'}}\rangle=\delta_{m,m'}\,\delta_{m_2,m'_2}
\no\\
&&\quad\quad\quad\quad\times \prod_{k=1}^{m}\delta_{p_k,q_k}\,
G_m(\theta_{p_1},\cdots,\theta_{p_{m_2}}|\theta_{p_{m_2+1}},\cdots,\theta_{p_m}),\label{Orthnormal}
\eea where the factor $G_m(\theta_{p_1},\cdots,\theta_{p_{m_2}}|\theta_{p_{m_2+1}},\cdots,\theta_{p_m})$ is given by
\bea
&&\hspace{-1.2truecm}G_m(\theta_{p_1},\cdots,\theta_{p_{m_2}}|\theta_{p_{m_2+1}},\cdots,\theta_{p_m})=\prod_{k=1}^{m_2}\sinh\eta\, d_{p_k}(\theta_{p_k})\,a(\theta_{p_k})\prod_{l=1,l\neq k}^{m_2}\frac{\sinh(\theta_{p_k}-\theta_{p_l}+\eta)}{\sinh(\theta_{p_k}-\theta_{p_l})}\no\\[4pt]
&&\quad\quad\times
\prod_{k=m_2+1}^m\sinh\eta \,d_{p_k}(\theta_{p_k})\,a(\theta_{p_k})\lt\{
\prod_{l=m_2+1,l\neq k}^m
\frac{\sinh(\theta_{p_k}-\theta_{p_l}+\eta)}{\sinh(\theta_{p_k}-\theta_{p_l})}
\rt.\no\\[4pt]
&&\quad\quad\quad\quad\times\lt.
\prod_{l=1}^{m_2}\frac{\sinh(\theta_{p_k}-\theta_{p_l}-\eta)}{\sinh(\theta_{p_k}-\theta_{p_l})}\rt\}.\label{Normal-factor}
\eea
Here the functions $\{d_l(u)\}$ are given by
\bea
d_l(u)=\prod_{k=1,k\neq l}^N\sinh(u-\theta_k),\quad l=1,\cdots,N. \label{d-function-1}
\eea
On the other hand, we know that the total number of the linear-independent left (right) states given in (\ref{left-state-su(3)}) ((\ref{right-state-su(3)})) is
\begin{eqnarray}
\sum_{m=0}^N\frac{N!}{(N-m)!m!}\sum_{m_2=0}^m\frac{m!}{(m-m_2)!m_2!}&=&\sum_{m=0}^N\frac{N!}{(N-m)!m!}\,2^m=3^N.
\end{eqnarray}
Thus these right (left) states
form an orthogonal right (left) basis of the Hilbert space, namely,
\bea
&&{\rm id}=\sum_{m=0}^N\sum_{m_2=0}^m\sum_{P}\frac{1}
{G_m(\theta_{p_1},\cdots,\theta_{p_{m_2}}|\theta_{p_{m_2+1}},\cdots,\theta_{p_m})}\no\\[4pt]
&&\quad\quad\quad\quad\times |
\theta_{p_1},\cdots,\theta_{p_{m_2}};\theta_{p_{m_2+1}},\cdots,\theta_{p_m}\rangle\,
\langle
\theta_{p_1},\cdots,\theta_{p_{m_2}};\theta_{p_{m_2+1}},\cdots,\theta_{p_m}|,\label{Identity}
\eea where the notation $\sum_{P}$ indicates the sum over all  $P$
satisfying the condition (\ref{P-condition}). Hence any right (left)
state can be decomposed as a unique linear combination of these
basis. Moreover, direct calculation shows that actions of the
monodromy matrix elements on this basis become drastically simple
(see below (\ref{D-decomposition})-(\ref{C3-decomposition})). Here
we list some of them relevant for us to construct eigenstates of the
transfer matrix in the next section, \bea &&\langle
\theta_{p_1},\cdots,\theta_{p_{m_2}};\theta_{p_{m_2+1}},\cdots,\theta_{p_m}|\,D^3_3(u)=d(u)\prod_{l=m_2+1}^{m}
\frac{\sinh(u-\theta_{p_{l}}+\eta)}
{\sinh(u-\theta_{p_{l}})}\no\\[4pt]
&&\qquad\qquad\qquad \quad \times \langle
\theta_{p_1},\cdots,\theta_{p_{m_2}};\theta_{p_{m_2+1}},\cdots,\theta_{p_m}|,\label{D-decomposition}\eea
\bea &&\langle
\theta_{p_1},\cdots,\theta_{p_{m_2}};\theta_{p_{m_2+1}},\cdots,\theta_{p_m}|\,D^2_3(u)=
\sum_{l=m_2+1}^m\frac{\sinh\eta\,e^{\frac{u-\theta_{p_l}}{3}}\,d(u)}{\sinh(u-\theta_{p_l})}\no\\[4pt]
&&\quad\qquad \quad \times \prod_{k=m_2+1,k\neq
l}^m\frac{\sinh(u-\theta_{p_k}+\eta)}{\sinh(u-\theta_{p_k})}
\frac{\sinh(\theta_{p_l}-\theta_{p_k}-\eta)}{\sinh(\theta_{p_l}-\theta_{p_k})}\no\\[4pt]
&&\quad\qquad \quad \times \langle
\theta_{p_1},\cdots,\theta_{p_{m_2}},\theta_{p_l};\theta_{p_{m_2+1}},\cdots,\theta_{p_{l-1}},\theta_{p_{l+1}},\cdots,\theta_{p_m}|,
\label{D23-decomposition}\eea \bea
 &&\langle
\theta_{p_1},\cdots,\theta_{p_{m_2}};\theta_{p_{m_2+1}},\cdots,\theta_{p_m}|\,D^3_2(u)=
\sum_{l=1}^{m_2}\frac{\sinh\eta\,e^{-\frac{u-\theta_{p_l}}{3}}\,
d(u)}{\sinh(u-\theta_{p_l})} \no \\[4pt]
&&\qquad \qquad\times \lt\{ \prod_{k=1,k\neq
l}^{m_2}\frac{\sinh(\theta_{p_l}-\theta_{p_k}+\eta)}{\sinh(\theta_{p_l}-\theta_{p_k})}
\prod_{k=m_2+1}^m\frac{\sinh(u-\theta_{p_k}+\eta)}{\sinh(u-\theta_{p_k})}\rt\} \no \\[4pt]
&&\qquad \qquad\times \langle
\theta_{p_1},\cdots,\theta_{p_{l-1}},\theta_{p_{l+1}},\cdots,\theta_{p_{m_2}};
\theta_{p_{m_2+1}},\cdots,\theta_{p_m},\theta_{p_l}|,\label{D32-decomposition}
\eea \bea &&\hspace{-1.6truecm}\langle
\theta_{p_1},\cdots,\theta_{p_{m_2}};\theta_{p_{m_2+1}},\cdots,\theta_{p_m}|\,B_3(u)=
\sum_{l=m_2+1}^m\frac{\sinh\eta\,e^{-\frac{u-\theta_{p_l}}{3}} d(u)}{\sinh(u-\theta_{p_l})}\,a(\theta_{p_l})\no\\[4pt]
&&\hspace{-0.6truecm}\quad\times\prod_{k=m_2+1,k\neq l}^m\frac{\sinh(u-\theta_{p_k}+\eta)}{\sinh(u-\theta_{p_k})} \frac{\sinh(\theta_{p_l}-\theta_{p_k}-\eta)}{\sinh(\theta_{p_l}-\theta_{p_k})}\no\\[4pt]
&&\hspace{-0.6truecm}\quad\times\prod_{\a=1}^{m_2}\frac{\sinh(\theta_{p_l}\hspace{-0.06truecm}-\hspace{-0.06truecm}
\theta_{p_{\a}}\hspace{-0.06truecm}-\hspace{-0.06truecm}\eta)}
{\sinh(\theta_{p_l}-\theta_{p_{\a}})}
\langle \theta_{p_1},\cdots,\theta_{p_{m_2}};\theta_{p_{m_2+1}},\cdots,\theta_{p_{l-1}},\theta_{p_{l+1}},\cdots,\theta_{p_m}|\no\\[4pt]
&&\hspace{-0.8truecm}+\sum_{l=m_2+1}^m\frac{\sinh\eta\,e^{-\frac{u-\theta_{p_l}}{3}} d(u)}{\sinh(u-\theta_{p_l})}
\prod_{k=m_2+1,k\neq l}^m\frac{\sinh(u\hspace{-0.06truecm}-\hspace{-0.06truecm}\theta_{p_k}\hspace{-0.06truecm}+\hspace{-0.06truecm}\eta)}
{\sinh(u-\theta_{p_k})}
\frac{\sinh(\theta_{p_l}\hspace{-0.06truecm}-\hspace{-0.06truecm}\theta_{p_k}\hspace{-0.06truecm}-\hspace{-0.06truecm}\eta)}
{\sinh(\theta_{p_l}-\theta_{p_k})}\no\\[4pt]
&&\hspace{-0.6truecm}\quad\times\sum_{\a=1}^{m_2}\frac{\sinh\eta\,e^{-\frac{\theta_{p_{\a}}-\theta_{p_l}}{3}}}
{\sinh(\theta_{p_l}-\theta_{p_{\a}})}
\,a(\theta_{p_{\a}})
\prod_{k=1,k\neq\a}^{m_2}\frac{\sinh(\theta_{p_{\a}}-\theta_{p_{k}}-\eta)}{\sinh(\theta_{p_{\a}}-\theta_{p_{k}})}
\no\\[4pt]
&&\hspace{-0.6truecm}\quad\times\langle \theta_{p_1},\cdots,\theta_{p_{\a-1}},\theta_{p_l},\theta_{p_{\a+1}},\cdots,\theta_{p_{m_2}};
\theta_{p_{m_2+1}},\cdots,\theta_{p_{l-1}},\theta_{p_{l+1}},\cdots,\theta_{p_m}|,\label{B3-decomposition}
\eea
\bea
&&\hspace{-1.6truecm}\langle \theta_{p_1},\cdots,\theta_{p_{m_2}};\theta_{p_{m_2+1}},\cdots,\theta_{p_m}|\,C^3(u)=
\sum_{l=m+1}^N\frac{e^{\frac{u-\theta_{p_l}}{3}}}{\sinh(u-\theta_{p_l})}\frac{d(u)}{d_{p_l}(\theta_{p_l})}\no\\[4pt]
&&\hspace{-0.6truecm}\quad\times\prod_{k=m_2+1}^m\frac{\sinh(u-\theta_{p_k}+\eta)}{\sinh(u-\theta_{p_k})}
\frac{\sinh(\theta_{p_l}-\theta_{p_k})}{\sinh(\theta_{p_l}-\theta_{p_k}+\eta)}
\no\\[4pt]
&&\hspace{-0.6truecm}\quad\times\langle \theta_{p_1},\cdots\theta_{p_{m_2}};
\theta_{p_{m_2+1}},\cdots,\theta_{p_{m}},\theta_{p_{l}}|\no\\[4pt]
&&\hspace{-0.8truecm}+\sum_{l=m+1}^N\sum_{\a=1}^{m_2}\frac{e^{\frac{u-\theta_{p_{\a}}}{3}}}{\sinh(u-\theta_{p_{\a}})}
\prod_{k=m_2+1}^m\frac{\sinh(u-\theta_{p_k}+\eta)}{\sinh(u-\theta_{p_k})}
\frac{\sinh(\theta_{p_l}-\theta_{p_k})}{\sinh(\theta_{p_l}-\theta_{p_k}+\eta)}\no\\[4pt]
&&\hspace{-0.6truecm}\quad\times \frac{\sinh\eta\,d(u)\,e^{\frac{\theta_{p_l}-\theta_{p_{\a}}}{3}}}
{d_{p_l}(\theta_{p_l})\sinh(\theta_{p_{\a}}\hspace{-0.06truecm}-\hspace{-0.06truecm}\theta_{p_l}\hspace{-0.06truecm}-\hspace{-0.06truecm}\eta)}
\prod_{k=1,k\neq\a}^{m_2}\frac{\sinh(\theta_{p_l}-\theta_{p_k})}
{\sinh(\theta_{p_l}\hspace{-0.06truecm}-\hspace{-0.06truecm}\theta_{p_k}\hspace{-0.06truecm}+\hspace{-0.06truecm}\eta)}
\frac{\sinh(\theta_{p_{\a}}\hspace{-0.06truecm}-\hspace{-0.06truecm}\theta_{p_k}\hspace{-0.06truecm}+\hspace{-0.06truecm}\eta)}
{\sinh(\theta_{p_{\a}}-\theta_{p_k})}\no\\[4pt]
&&\hspace{-0.6truecm}\quad\times\langle \theta_{p_1},\cdots,\theta_{p_{\a-1}},\theta_{p_l},\theta_{p_{\a+1}},\cdots,\theta_{p_{m_2}};
\theta_{p_{m_2+1}},\cdots,\theta_{p_{m}},\theta_{p_{\a}}|.\label{C3-decomposition}
\eea
The sketch proof of the above operator decompositions is given in Appendix C. Similarly, one may derive operator decompositions on the right basis which also have simple forms as (\ref{D-decomposition})-(\ref{C3-decomposition}).

Some remarks are in order. In the rational limit\footnote{Redefine: $u\rightarrow \e u$, $\theta_j\rightarrow \e\theta_j$ and $\eta\rightarrow
\e \eta$, then take the limit  $\e\rightarrow 0$. }, the resulting
basis serves as the SoV basis for the associated rational spin chain model\footnote{The resulting SoV basis for the rational spin chain model is different from that in \cite{Skl92-1}. It is
interesting to study the relation between them.}.
We have checked that each basis vector given by (\ref{left-state-su(3)}) and (\ref{right-state-su(3)})
for the $su(3)$ case (the generalizations to the $su(n)$ case are given in Appendix B, see   (\ref{left-basis}) and (\ref{right-basis}) below ) is an off-shell Bethe state obtained
via the nested algebraic Bethe Ansatz \cite{2sun1} by replacing the Bethe roots with some sets of the inhomogeneity parameters\footnote{A general off-shell Bethe state is  $\overline{|\l_1,\cdots,\l_m;\l^{(1)}_1,\cdots,\l^{(1)}_{m-m_2}\rangle}=B_{i_1}(\l_1)\cdots B_{i_m}(\l_m)\,F^{i_1,\cdots,i_m}|0\rangle$, where $\{F^{i_1,\cdots,i_m}|i_l=2,3\}$ are the vector components of a nested off-shell Bethe state $B^{(1)}(\l^{(1)}_1)\cdots B^{(1)}(\l^{(1)}_{m-m_2})|0\rangle^{(1)}=\sum_{i_1,\cdots,i_m=2}^3F^{i_1,\cdots,i_m}|i_1,\cdots,i_m\rangle^{(1)}$,
and the operator $B^{(1)}(u)$ and $|0\rangle^{(1)}$ are the corresponding creation operator and the reference state associated with the nested $su(2)$ spin chain with $m$ sites and the corresponding inhomogeneous parameters being $\{\l_1,\cdots,\l_m\}$ \cite{2sun1}. For general values of $\l_1,\cdots,\l_m$ and $\l^{(1)}_1,\cdots,\l^{(1)}_{m-m_2}$, the Bethe sate
$\overline{|\l_1,\cdots,\l_m;\l^{(1)}_1,\cdots,\l^{(1)}_{m-m_2}\rangle}$ is a linear combination of the vectors (\ref{right-state-su(3)}) \cite{Bel13-1}. However, if the parameters $\{\l_l|l=1,\cdots,m\}$ are particularly chosen as $\{\l_l=\theta_{p_l}|l=1,\cdots,m\}$ and then the nested parameters $\{\l^{(1)}_n|n=1,\cdots,m-m_2\}$ have to take the values in the chosen set of $\{\l_l|l=1,\cdots,m\}$ (e.g., $\{\l^{(1)}_n=\theta_{p_n}|n=m_2+1,\cdots,m\}$), the corresponding linear combination becomes drastically simple such that  only one term such as (\ref{right-state-su(3)}) does remain. }. This observation provides
an efficient way to construct similar nested SoV basis for general high-rank quantum integrable models.
From explicit expressions (\ref{D-decomposition})-(\ref{C3-decomposition}),
one can see that in the basis (\ref{left-state-su(3)}) the operators have no compensating exchange terms on the level of the local operators (i.e.
polarization free), which have similar simple forms as those in the F-basis \cite{Mai96,Alb00, Alb01, Yan06} and allow us to compute correlation functions \cite{Kor93} for  quantum
spin chains associated with  higher-rank algebras \cite{Zha06,Bel13-1}.

\section{Eigenstates of the transfer matrix}
\setcounter{equation}{0}
In this section, we adopt the method developed in \cite{Cao-14-Bethe-state} (see  also \cite{Wan15}) to construct eigenstates of the $su(3)$ spin torus based on the inhomogeneous
$T-Q$ relations given by (\ref{T-Q-1}) \cite{Hao16} and the basis introduced in the previous section.
For the $su(3)$ case, the monodromy matrix is expressed in terms of the operators (\ref{A-operator})-(\ref{D-operator}) as
\bea
T(u)=\lt(\begin{array}{ccc}A(u)&B_2(u)&B_3(u)\\
C^2(u)&D^2_2(u)&D^2_3(u)\\
C^3(u)&D^3_2(u)&D^3_3(u)
\end{array} \rt).
\eea The corresponding transfer matrix (\ref{transfer}) reads \bea
t(u)=B_2(u)+D^2_3(u)+C^3(u).\label{Transfer-3} \eea The
commutativity of the transfer matrices $t(u)$  with different
spectral parameters implies that they have common eigenstates. Let
$|\Psi\rangle$ be a common eigenstate of $t(u)$, which does not
depend upon $u$, with an eigenvalue $\Lambda(u)$, i.e., \bea
t(u)|\Psi\rangle=\Lambda(u)|\Psi\rangle,\no \eea where the
eigenvalue $\L(u)$ of the transfer matrix $t(u)$ is given by the
inhomogeneous $T-Q$ relation (\ref{T-Q-1}). Due to the fact that the
left states $\{\langle
\theta_{p_1},\cdots,\theta_{p_{m_2}};\theta_{p_{m_2+1}},\cdots,\theta_{p_m}|,
m_2=0,\cdots,m; \,m=0,\cdots, N\}$ given by (\ref{left-state-su(3)})
form a basis of the dual Hilbert space, the eigenstate
$|\Psi\rangle$ is completely determined (up to an overall scalar
factor) by the following scalar products
\cite{Cao1,Cao-14-Bethe-state}
\begin{eqnarray}
&&\hspace{-1.2truecm}F_{m_2,m-m_2}(\theta_{p_1},\cdots,\theta_{p_{m_2}};\theta_{p_{m_2+1}},\cdots,\theta_{p_m})=
\langle \theta_{p_1},\cdots,\theta_{p_{m_2}};\theta_{p_{m_2+1}},\cdots,\theta_{p_m}|\Psi\rangle,\no\\[4pt]
&&\hspace{-1.2truecm}\quad  1\leq p_1<\cdots<p_{m_2}, \, 1\leq
p_{m_2+1}<\cdots<p_{m}\leq N,\,p_j\neq p_k, \,\, 0\leq m_2\leq m\leq
N. \label{Scalar-product-1}
\end{eqnarray} Following \cite{Cao-14-Bethe-state}, let us consider the quantities
$\langle \theta_{p_1},\cdots,\theta_{p_{m_2}};\theta_{p_{m_2+1}},\cdots,\theta_{p_m}|t(\theta_{p_{m+1}})|\Psi\rangle$.
Acting $t(\theta_{p_{m+1}})$ to the right gives rise to the relation
\bea
&&\Lambda(\theta_{p_{m+1}})\,F_{m_2,m-m_2}(\theta_{p_1},\cdots,\theta_{p_{m_2}};\theta_{p_{m_2+1}},\cdots,\theta_{p_m})\no\\[4pt]
&&\quad\quad\quad\quad= \langle
\theta_{p_1},\cdots,\theta_{p_{m_2}};\theta_{p_{m_2+1}},\cdots,\theta_{p_m}|t(\theta_{p_{m+1}})|\Psi\rangle.
\eea With the help of  the expression (\ref{Transfer-3}) of the
transfer matrix and the relations
(\ref{Vanshing-1})-(\ref{Vanshing-2}), by acting
$t(\theta_{p_{m+1}})$ to the left we readily obtain \bea
F_{m_2,m-m_2}(\theta_{p_1},\cdots,\theta_{p_{m_2}};\theta_{p_{m_2+1}},\cdots,\theta_{p_m})=
\lt\{\prod_{l=m_2+1}^m\L(\theta_{p_{l}})\rt\}\,F_{m_2}(\theta_{p_{1}},\cdots,\theta_{p_{m_2}}),
\label{Scalar-product-2} \eea where the scalar products
$F_{m}(\theta_{p_1},\cdots,\theta_{p_{m}})$ are given by \bea
F_{m}(\theta_{p_1},\cdots,\theta_{p_{m}})=\langle
0|C^2(\theta_{p_1})\cdots C^2(\theta_{p_{m}})|\Psi\rangle,\quad
m=0,\cdots,N. \label{Scalar-product-3} \eea It follows that in order
to obtain all the scalar products (\ref{Scalar-product-1}) it is
sufficient to compute the scalar products (\ref{Scalar-product-3}).
After a tedious calculation, we have \bea
&&\hspace{-1.2truecm}F_m(\theta_{p_1},\cdots,\theta_{p_m})
=\sum_{1\leq p'_1<\cdots< p'_m\leq
N}\,g_m(\theta_{p_1},\cdots,\theta_{p_m}|\theta_{p'_1},\cdots,\theta_{p'_m})
\no\\[4pt]
&&\quad
\times\prod_{\a=1}^m\prod_{k=m+1}^N\sinh(\theta_{p'_{\a}}-\theta_{p_k}+\eta)
\frac{\prod_{l=1}^m\Lambda(\theta_{p'_l})}{f_m(\theta_{p'_1},\cdots,\theta_{p'_m})}
\frac{\prod_{l=1}^Na(\theta_k)}{\prod_{k=m+1}^N\Lambda(\theta_{p_k})}\langle\bar{0}|\Psi\rangle,\label{Scalar-product-4}
\eea where the state $\langle\bar{0}|=\langle 3,\cdots,3|$ and the
functions $g_m(v_1,\cdots,v_m|u_1,\cdots,u_m)$ and
$f_m(\theta_{p_1}, $ $\cdots, \theta_{p_m})$ are given by \bea
&&\hspace{-1.2truecm}g_m(v_1,\cdots,v_m|u_1,\cdots,u_m)=\frac{\prod_{\a=1}^m\prod_{k=1}^m\sinh(u_{\a}-v_k+\eta)\sinh(u_{\a}-v_k)}
{\prod_{k<l}^m\sinh(u_l-u_k)\sinh(v_k-v_l)}
\det{\cal {M}}, \label{g-function}\\[4pt]
&&\hspace{-1.2truecm}f_m(\theta_{p_1},\cdots,\theta_{p_m})=\prod_{l=1}^m\sinh\eta \,d_{p_l}(\theta_{p_l})\,a(\theta_{p_l})
\prod_{k=1,k\neq l}^m
\frac{\sinh(\theta_{p_l}-\theta_{p_k}+\eta)}{\sinh(\theta_{p_l}-\theta_{p_k})},\label{f-function}
\eea and ${\cal{M}}$ is an $m\times m$ matrix  with  matrix elements
\bea
{\cal{M}}_{\a,k}=\frac{\sinh\eta\,e^{-\frac{u_{\a}-v_k}{3}}}{\sinh(u_{\a}-v_k+\eta)\,\sinh(u_{\a}-v_k)},\quad \a,k=1,\cdots,m.
\eea
The proof of (\ref{Scalar-product-4}) is given in Appendix D.

The identity decomposition (\ref{Identity}) allows us to retrieve the eigenstate $|\Psi\rangle$
of the transfer matrix corresponding to an eigenvalue $\Lambda(u)$ as
\bea
|\Psi\rangle&=&
\sum_{m=0}^N\sum_{m_2=0}^m\sum_{P}
\frac{\langle \theta_{p_1},\cdots,\theta_{p_{m_2}};\theta_{p_{m_2+1}},\cdots,\theta_{p_m}|\Psi\rangle}
{G_m(\theta_{p_1},\cdots,\theta_{p_{m_2}}|\theta_{p_{m_2+1}},\cdots,\theta_{p_m})}\no\\[4pt]
&&\quad\times  | \theta_{p_1},\cdots,\theta_{p_{m_2}};\theta_{p_{m_2+1}},\cdots,\theta_{p_m}\rangle\no\\[4pt]
&=&\sum_{m=0}^N\sum_{m_2=0}^m\sum_{P}
\frac{F_{m_2}(\theta_{p_1},\cdots,\theta_{p_{m_2}})\prod_{k=m_2+1}^m\Lambda(\theta_{p_k})}
{G_m(\theta_{p_1},\cdots,\theta_{p_{m_2}}|\theta_{p_{m_2+1}},\cdots,\theta_{p_m})}\no\\[4pt]
&&\quad\times  | \theta_{p_1},\cdots,\theta_{p_{m_2}};\theta_{p_{m_2+1}},\cdots,\theta_{p_m}\rangle,\label{Eigen-state-decomposition}
\eea
where the factors $F_{m_2}(\theta_{p_1},\cdots,\theta_{p_{m_2}})$  and $G_m(\theta_{p_1},\cdots,\theta_{p_{m_2}}|\theta_{p_{m_2+1}},\cdots,\theta_{p_m})$
are given respectively by (\ref{Scalar-product-4}) and (\ref{Normal-factor}). It should be emphasized that the factor $F_{m_2}(\theta_{p_1},\cdots,\theta_{p_{m_2}})$
does depend upon the corresponding eigenvalue $\Lambda(u)$ associated with the eigenstate $|\Psi\rangle$, while $G_m(\theta_{p_1},\cdots,\theta_{p_{m_2}}|\theta_{p_{m_2+1}},\cdots,\theta_{p_m})$ does not.

Some remarks are in order. In the homogeneous limit, the resulting eigenstate (\ref{Eigen-state-decomposition}) (if it exists) becomes
the eigenstate of the homogeneous quantum spin chain (i.e. the $su(3)$ spin torus) due to the fact the $T-Q$ relation (\ref{T-Q-1}) and the associated
BAEs (\ref{BAE-1})-(\ref{BAE-8}) have well-defined homogeneous limits.  We have checked that such a limit of the state (\ref{Eigen-state-decomposition})
does exist for some small $N$.  For an example, here we present the limit of the $N=2$ case
\bea
\lim_{\theta_1,\theta_2\rightarrow 0}|\Psi\rangle&\propto&|0\rangle
+\frac{1}{\sinh^3\eta}\lt[\L'\, B_3+\L\, B'_3-2\coth\eta\, \L\, B_3\rt]|0\rangle
+\frac{\L^2}{\sinh^8\eta}B_3\, B_3|0\rangle\no\\[6pt]
&&+\frac{\L^2}{a^2(0)}\lt\{\lt[\lt(\frac{8}{9}-2\coth\eta\,\frac{\L'}{\L}+(\frac{\L'}{\L})^2\rt)B_2+
         \lt(\frac{\L'}{\L}-\coth\eta-\frac{1}{3}\rt)B'_2\rt]\rt.\no\\[6pt]
&&\quad\quad\quad\quad +\frac{\L}{\sinh^4\eta}\lt[\lt(\coth\eta \frac{\L'}{\L}-\frac{\L'}{3\L}-\frac{8}{9}\rt)B_3\,B_2\rt.\no\\[6pt]
&&\quad\quad\quad\quad\quad\quad\quad+\lt(\lt.\frac{\L'}{\L}-\coth\eta-\frac{1}{3}\rt)\lt(B'_3\,B_2-B_3\,B'_2\rt)\rt]\no\\[6pt]
&&\quad\quad\quad\quad+\lt.\frac{\L^2}{\sinh^8\eta}\,B_2\,B_2\rt\}|0\rangle,
\eea
where
\bea
&&B_i=B_i(0),\quad B'_i=\lt.\frac{\partial}{\partial u}\,B_i(u)\rt|_{u=0},\quad i=2,3,\no\\[6pt]
&&\L=\L(0),\quad \L'=\lt.\frac{\partial}{\partial u}\,\L(u)\rt|_{u=0}.\no
\eea
It is  conjectured that  the eigenstate (\ref{Eigen-state-decomposition}) for generic $N$ has a well-defined homogeneous limit. However, the direct proof remains
an important open problem.


\section{Conclusions}
In this paper, we introduced a convenient basis of the Hilbert
space, which could be treated as the SoV basis for the trigonometric
spin chain model associated with the $su(3)$ algebra. We have demonstrated that the
monodromy matrix elements acting on a generic basis vector take
simple forms such as
(\ref{D-decomposition})-(\ref{C3-decomposition}) without
compensating exchange terms on the level of the local operators
(i.e. polarization free). With the help of this basis, the
corresponding eigenstates of the transfer matrix can be constructed
by (\ref{Eigen-state-decomposition}) via its ODBA solution
\cite{Hao16}. In the rational limit, the resulting basis serves as
the SoV basis for the associated rational spin chain model.
Moreover, as each basis vector is an off-shell Bethe state with the
Bethe roots replaced by the inhomogeneous parameters, this
procedure provides an efficient way to construct nested SoV basis
for generic high-rank quantum integrable models such as the one-dimensional Hubbard model \cite{mar} with algebraic Bethe
Ansatz.

\section*{Acknowledgments}

The financial supports from the National Natural Science Foundation
of China (Grant Nos. 11375141, 11374334, 11434013, 11425522 and 11547045), BCMIIS and the Strategic Priority Research Program
of the Chinese Academy of Sciences are gratefully acknowledged.

\section*{Appendix A: Exchange relations}
\setcounter{equation}{0}
\renewcommand{\theequation}{A.\arabic{equation}}
The QYBE implies the following exchange relations among the monodromy matrix elements
\bea
C^l(v)\,D^k_i(u)&=&\sum_{\a,\b=2}^n\frac{{R}^{k\,l}_{\a\,\b}(u-v)}{\sinh(u-v)}\,D^{\a}_i(u)\,C^{\b}(v)-\frac{R^{1\,i}_{i\,1}(u-v)}{\sinh(u-v)}\,
D^l_i(v)\,C^k(u),\label{CD}\\[6pt]
C^{k}(v)\,A(u)&=&\frac{\sinh(u-v-\eta)}{\sinh(u-v)}A(u)C^k(v)+\frac{R^{k\,1}_{1\,k}(v-u)}{\sinh(u-v)}A(v)C^k(u),\label{CA}
\eea
\bea
[C^i(u),\,B_l(v)]&=&\frac{1}{\sinh(u-v)}\,\lt(R^{l\,1}_{1\,l}(u-v)A(v)\,D^i_l(u)-R^{i\,1}_{1\,i}(u-v)A(u)\,D^i_l(v)\rt)\no\\[6pt]
&=&\frac{1}{\sinh(u-v)}\,\lt(R^{1\,l}_{l\,1}(v-u)D^i_l(u)\,A(v)-R^{1\,i}_{i\,1}(v-u)D^i_l(v)\,A(u)\rt),\label{CB}
\eea
\bea
A(u)\,B_i(v)&=&\frac{\sinh(u-v-\eta)}{\sinh(u-v)} B_i(v)\,A(u)+\frac{R^{1\,i}_{i\,1}(v-u)}{\sinh(u-v)} B_i(u)\,A(v),\label{AB}\\[6pt]
D^j_i(u)\, B_l(v)&=&\sum_{\a,\b=2}^n\frac{R^{\a\,\b}_{i\,\,\,l}(u-v)}{\sinh(u-v)} \,B_{\b}(v) \,D^j_{\a}(u)-\frac{R^{j1}_{1j}(u-v)}{\sinh(u-v)} B_i(u)\, D^j_l(v),\label{DB}\\[6pt]
B_i(u)\,B_j(v)&=&\sum^n_{\alpha,\beta=2}\frac{R^{\a\,\b}_{i\,\,j}(u-v)}{\sinh(u-v+\eta)}\, B_{\b}(v)\,B_{\a}(u),\label{BB}\\[6pt]
C^{j}(v)\,C^{i}(u)&=&\sum_{\a,\b=2}^n\frac{{R}^{i\,j}_{\a\,\b}(u-v)}{\sinh(u-v+\eta)}\,C^{\a}(u)\,C^{\b}(v),\label{CC}
\eea
\bea
&&[T^{\a}_{\b}(u),\,T^{\a}_{\b}(v)]=0,\quad \a,\,\b=1,\cdots,n,\label{TT-1}\\[6pt]
&&[T^{\a}_{\a}(u),\, T^{\b}_{\b}(v)]=\frac{1}{\sinh(u-v)}\lt\{
R^{\b\,\a}_{\a\,\b}(u-v)T^{\b}_{\a}(v)\,T^{\a}_{\b}(u) \rt. \no \\[6pt]
&& \qquad  \qquad  \qquad  \quad \lt. -R^{\a\,\b}_{\b\,\a}(u-v)T^{\b}_{\a}(u)\,T^{\a}_{\b}(v)\rt\}, \quad \a\neq \b,\,\,{\rm and}\,\,\a,\,\b=1,\cdots,n,\label{TT-2}\\[6pt]
&&[T^{\a}_{\b}(u),\, T^{\b}_{\a}(v)]=\frac{R^{\a\,\b}_{\b\,\a}(u-v)}
{\sinh(u-v)}\lt\{T^{\b}_{\b}(v)\,T^{\a}_{\a}(u)-T^{\b}_{\b}(u)\,T^{\a}_{\a}(v)\rt\},\no\\[6pt]
&&\qquad  \quad \qquad  \quad \qquad  \a\neq\b,\,\,{\rm
and}\,\,\a,\,\b=1,\cdots,n.\label{TT-3} \eea

\section*{Appendix B: $su(n)$ case}
\setcounter{equation}{0}
\renewcommand{\theequation}{B.\arabic{equation}}
For the $su(n)$ spin chain, let us introduce $n-1$ non-negative
integers $m_2,m_3,\cdots,m_n$ such that $\sum_{l=2}^nm_l\leq N$ and
\bea && \langle \theta_{p_1},\cdots,\theta_{p_{m_2}};\cdots
;\theta_{p_{m_2+\cdots m_{n-1}+1}},\cdots, \theta_{p_{m_2+\cdots m_{n}}}|=\langle 0|\,C^2(\theta_{p_1})\cdots C^2(\theta_{p_{m_2}})\cdots\no\\[6pt]
&&\quad \quad \quad \quad \times C^{n}(\theta_{p_{m_2+\cdots+m_{n-1}+1}})\cdots C^{n}(\theta_{p_{m_2+\cdots+m_{n}}}),\label{left-basis}\\[6pt]
&& |\theta_{p_1},\cdots,\theta_{p_{m_2}};\cdots
;\theta_{p_{m_2+\cdots m_{n-1}+1}},\cdots, \theta_{p_{m_2+\cdots m_{n}}}\rangle=B_{n}(\theta_{p_{m_2+\cdots+m_{n}}})\cdots
\no\\[6pt]
&&\quad\quad\quad\quad\times
B_{n}(\theta_{p_{m_2+\cdots+m_{n-1}+1}})\cdots
B_2(\theta_{p_{m_2}})\cdots B_2(\theta_{p_1})\,|0\rangle,
\label{right-basis} \eea where $1\leq p_1<\cdots<p_{m_2}\leq N$,
$\cdots$, $1\leq
p_{m_2+\cdots+m_{n-1}+1}<\cdots<p_{m_2+\cdots+m_{n}}\leq N$ and
$p_j\neq p_k$. Note that the number of the operators $C^j(u)$ (or
$B_j(u)$) in the above expression  is $m_j$. Due to the fact that
$d(\theta_j)=0$, with the help of the exchange relations (\ref{CD})
and (\ref{DB}), we can show that these states are in fact
eigenstates of the operator $D^n_n(u)$
\bea &&\langle
\theta_{p_1},\cdots,\theta_{p_{m_2}};\cdots ;\theta_{p_{m_2+\cdots
m_{n-1}+1}},\cdots, \theta_{p_{m_2+\cdots m_{n}}}|\,D^n_n(u)\no\\[6pt]
&&\qquad =
 d(u)\prod_{k=m_2+\cdots m_{n-1}+1}^{m_2+\cdots m_n}
\frac{\sinh(u-\theta_{p_k}+\eta)}{\sinh(u-\theta_{p_k})}\no\\[6pt]
&&\quad\quad\quad\quad \times
 \langle \theta_{p_1},\cdots,\theta_{p_{m_2}};\cdots
;\theta_{p_{m_2+\cdots m_{n-1}+1}},\cdots, \theta_{p_{m_2+\cdots m_{n}}}|,\label{eigenvalue-left}\\[6pt]
&&D^n_n(u)\,|\theta_{p_1},\cdots,\theta_{p_{m_2}};\cdots
;\theta_{p_{m_2+\cdots m_{n-1}+1}},\cdots, \theta_{p_{m_2+\cdots
m_{n}}}\rangle \no \\[6pt]
&&\qquad=
 d(u)\prod_{k=m_2+\cdots m_{n-1}+1}^{m_2+\cdots m_n}
\frac{\sinh(u-\theta_{p_k}+\eta)}{\sinh(u-\theta_{p_k})}\no\\[6pt]
&&\quad\quad\quad\quad \times |\theta_{p_1},\cdots,\theta_{p_{m_2}};\cdots
;\theta_{p_{m_2+\cdots m_{n-1}+1}},\cdots, \theta_{p_{m_2+\cdots m_{n}}}\rangle.
\label{eigenvalue-right}
\eea For generic values of $\{\theta_j\}$,  these right (left) states
form an orthogonal right (left) basis of the Hilbert space,
and any right (left) state can be decomposed as a unique
linear combination of these basis.

Using the similar method in Appendix C, we can obtain the explicit expressions for the operators $\{D^n_i(u), D_n^i(u)|i=2,\cdots,n\}$,
$B_n(u)$ and $C^n(u)$ in the basis (\ref{left-basis}). Like  (\ref{D-decomposition})-(\ref{C3-decomposition}), the operators take some
simple forms without compensating exchange terms on the level of the local operators (i.e. polarization free) and hence
have similar simple forms as those in the F-basis \cite{Mai96,Alb00, Yan06}. These resulting simple forms allow one to construct
eigenstates of the transfer matrix  of the $su(n)$ spin torus via its ODBA solution \cite{Hao16}.

\section*{Appendix C: Proof of the operator decomposition}
\setcounter{equation}{0}
\renewcommand{\theequation}{C.\arabic{equation}}
Keeping the relations (\ref{Eigenvalue-left-3}) and  (\ref{Vanshing-1})-(\ref{Vanshing-2}) in mind and using the exchange
relations (\ref{CD})-(\ref{TT-3}), one can easily check the actions (\ref{D-decomposition})-(\ref{B3-decomposition}) straightforwardly. In order
to prove (\ref{C3-decomposition}), we apply the identity decomposition (\ref{Identity}) to the LHS of (\ref{C3-decomposition}), giving rise to
\bea
&&\langle \theta_{p_1},\cdots,\theta_{p_{m_2}};\theta_{p_{m_2+1}},\cdots,\theta_{p_{m}}|C^3(u)=
\sum_{P'}\langle \theta_{p'_1},\cdots,\theta_{p'_{m_2}};\theta_{p'_{m_2+1}},\cdots,\theta_{p'_{m+1}}|\no\\[6pt]
&&\quad\quad\times \frac{\langle \theta_{p_1},\cdots,\theta_{p_{m_2}};\theta_{p_{m_2+1}},\cdots,\theta_{p_{m}}|C^3(u)| \theta_{p'_1},\cdots,\theta_{p'_{m_2}};\theta_{p'_{m_2+1}},\cdots,\theta_{p'_{m+1}}\rangle}
{G_{m+1}(\theta_{p'_1},\cdots,\theta_{p'_{m_2}}|\theta_{p'_{m_2+1}},\cdots,\theta_{p'_{m+1}})},
\eea
where the sum is over $P'=\{p'_1,\cdots,p'_{m+1}\}$ such that $1\leq p'_1<\cdots<p'_{m_2}\leq N$, $1\leq p'_{m2+1}<\cdots<p'_{m+1}\leq N$
and $p'_j\neq p'_k$.  Since that the factor $G_{m+1}(\theta_{p'_1},\cdots,\theta_{p'_{m_2}}|\theta_{p'_{m_2+1}},\cdots,\theta_{p'_{m+1}})$ is
already known (\ref{Normal-factor}), it is sufficient to compute the scalar products
\bea
&&\hspace{-0.8truecm}\langle \theta_{p_1},\cdots,\theta_{p_{m_2}};\theta_{p_{m_2+1}},\cdots,\theta_{p_{m}}|C^3(u)| \theta_{p'_1},\cdots,\theta_{p'_{m_2}};\theta_{p'_{m_2+1}},\cdots,\theta_{p'_{m+1}}\rangle\no\\[6pt]
&&\quad =\langle \theta_{p_1},\cdots,\theta_{p_{m_2}};\theta_{p_{m_2+1}},\cdots,\theta_{p_{m}}|\,C^3(u)\,
B_3(\theta_{p'_{m+1}})\cdots B_3(\theta_{p'_{m_2+1}})\no\\[6pt]
&&\quad\quad\quad\quad\quad\quad \times B_2(\theta_{p'_{m_2}})\cdots B_2(\theta_{p'_1})|0\rangle.
\eea
The relations (\ref{Eigenvalue-right-3}), (\ref{Vanshing-3})-(\ref{Vanshing-4}) and the exchange relations (\ref{CD})-(\ref{TT-3})
allow us to arrive at  the operator decomposition (\ref{C3-decomposition}) by direct calculation. Similarly, we can work out the
explicit decomposition expressions for the operators $D^2_2(u)$, $B_2(u)$, $C^2(u)$ and $A(u)$.

\section*{Appendix D: Proof of (\ref{Scalar-product-4})}
\setcounter{equation}{0}
\renewcommand{\theequation}{D.\arabic{equation}}
Let us introduce a subspace ${\cal{H}}_{m}$ for a fixed non-negative
integer $m$ spanned by the states \bea \langle
0|C^2(\theta_{p_1})\cdots
C^2(\theta_{p_{m}})\,C^3(\theta_{p_{m+1}})\cdots
C^3(\theta_{p_{N}}), \label{Hilbert-1} \eea where $1\leq
p_1<\cdots<p_{m}\leq N$, $1\leq p_{m+1}<\cdots<p_N\leq N$ and
$p_j\neq p_k$. It is easy to check that the dimension of the
subspace is $\frac{N!}{m!(N-m)!}$ and that the subspace can also be
spanned by another set of states \bea \langle
\bar{0}|\,D^2_3(\theta_{p_1})\cdots D^2_3(\theta_{p_{m}}),\quad
\langle \bar{0}|=\langle 3,\cdots,3|, \eea where $1\leq
p_1<\cdots<p_{m}\leq N$. Similar to the procedure for deriving
(\ref{Orthnormal}), we have \bea \langle
\bar{0}|\,D^2_3(\theta_{p_1})\cdots D^2_3(\theta_{p_{m}})
D^3_2(\theta_{q_{1}})\cdots D^3_2(\theta_{q_{m'}})|\bar{0}\rangle
=\delta_{m,m'}\prod_{k=1}^m\delta_{p_k,q_k}\,f_m(\theta_{p_1},\cdots,\theta_{p_m}),\label{D-3}
\eea where the state $|\bar{0}\rangle=|3,\cdots,3\rangle$ and the
normalized factor $f_m(\theta_{p_1},\cdots,\theta_{p_m})$ is given
by (\ref{f-function}). The relations
(\ref{Vanshing-1})-(\ref{Vanshing-2}) and the operator decomposition
(\ref{D32-decomposition}) enable us to derive that
\bea &&\langle
0|C^2(\theta_{p_1})\cdots
C^2(\theta_{p_{m}})\,C^3(\theta_{p_{m+1}})\cdots C^3(\theta_{p_{N}})
\,D^3_2(u_1)\cdots D^3_2(u_m)|\bar{0}\rangle\no\\[6pt]
&&\quad=
\prod_{\a=1}^m\prod_{k=m+1}^N\sinh(u_{\a}-\theta_{p_k}+\eta)\,g_m(\theta_{p_1},\cdots,\theta_{p_m}|u_1,\cdots,u_m)\langle
0|C^3(\theta_1)\cdots C^3(\theta_N)
|\bar{0}\rangle\no\\[6pt]
&&\quad=\prod_{\a=1}^m\prod_{k=m+1}^N\sinh(u_{\a}-\theta_{p_k}+\eta)\,g_m(\theta_{p_1},\cdots,\theta_{p_m}|u_1,\cdots,u_m)\,\prod_{k=1}^Na(\theta_k)\label{D-4},
\eea where the function $g_m(v_1,\cdots,v_m|u_1,\cdots,u_m)$ is
given by (\ref{g-function}) and we have used the identity: $\langle
0|C^3(\theta_1)\cdots
C^3(\theta_N)|\bar{0}\rangle=\prod_{k=1}^Na(\theta_k)$.

The corresponding matrix $g$ given by (\ref{g-matrix-3}) allows us
to introduce an operator $U(g)$ acting on the Hilbert space as
\bea
U(g)=g_1\,g_2\cdots\,g_N,\quad \lt\{U(g)\rt\}^3={\rm id}. \label{Quantum-g}
\eea
The invariant property (\ref{Invariant-R}) of the $R$-matrix and the definition (\ref{Monodromy-1}) of the monodromy matrix $T_0(u)$
give rise to  the relation
\bea
g_0\,T_0(u)\,g_0^{-1}=U^{-1}(g)\,T_0(u)\,U(g),
\eea
which implies that
\bea
U^{-1}(g)\,C^3(u)\,U(g)=D^2_3(u), \quad  U^{-1}(g)\,t(u)\,U(g)=t(u).\label{Invariant-transfer}
\eea
Direct calculation shows that
\bea
\langle 0|U(g)=\langle \bar{0}|. \label{State-bar}
\eea
The invariance (\ref{Invariant-transfer}) of the transfer matrix leads to that the state $U(g)\,|\Psi\rangle$ is also
an eigenstate of the transfer matrix with the same eigenvalue, namely,
\bea
t(u)\,U(g)\,|\Psi\rangle=\Lambda(u)\,U(g)\,|\Psi\rangle. \label{New-eigenstate}
\eea
Hence we can  simultaneously diagonalize the transfer matrix and the operator $U(g)$, i.e.,
\bea
U(g)\,|\Psi\rangle=\omega^{Z(|\Psi\rangle)}\,|\Psi\rangle,\quad  Z(|\Psi\rangle)=0,1,2.\label{Z-charge}
\eea
Moreover, with the help of the relations (\ref{Initial}) and (\ref{Unitarity}) we can show that
\bea
\prod_{j=1}^Nt(\theta_j)=\lt\{\prod_{j=1}^Na(\theta_j)\rt\}\times U(g),
\eea
which gives rise to the identity
\bea
\frac{\prod_{j=1}^N\Lambda(\theta_j)}{\prod_{j=1}^Na(\theta_j)}=\omega^{Z(|\Psi\rangle)}.\label{Z-charge-1}
\eea
The above identity allows one to compute the eigenvalue of the operator $U(g)$ for an eigenstate $|\Psi\rangle$ from the
associated Bethe ansatz solution given by (\ref{T-Q-1})-(\ref{BAE-8}).

The relations (\ref{Invariant-transfer})-(\ref{New-eigenstate})  allow us to derive that
\bea
\langle \bar{0}|D^2_3(\theta_{p_1})\cdots D^2_3(\theta_{p_m})|\Psi\rangle&=&\langle 0|
C^3(\theta_{p_1})\cdots C^3(\theta_{p_m})\,U(g)\,|\Psi\rangle\no\\[6pt]
&\stackrel{(\ref{Scalar-product-2})}{=}& \prod_{l=1}^m\Lambda(\theta_{p_l})\langle 0|U(g)|\Psi\rangle\no\\[6pt]
&\stackrel{(\ref{Z-charge})}{=}&\omega^{Z(|\Psi\rangle)}\,\prod_{l=1}^m\Lambda(\theta_{p_l})\langle 0|\Psi\rangle\no\\[6pt]
&\stackrel{(\ref{Z-charge-1})}{=}&\frac{\prod_{j=1}^N\Lambda(\theta_j)}{\prod_{j=1}^Na(\theta_j)}\times
\prod_{l=1}^m\Lambda(\theta_{p_l})\langle 0|\Psi\rangle.\label{D-5}
\eea

Now we are in position to prove (\ref{Scalar-product-4}). The relation (\ref{Scalar-product-2}) implies that
\bea
&&\hspace{-1.2truecm}F_{m}(\theta_{p_1},\cdots,\theta_{p_{m}})=\langle 0|C^2(\theta_{p_1})\cdots C^2(\theta_{p_m})|\Psi\rangle\no\\[6pt]
&&\hspace{-1.2truecm}\quad\quad=
\frac{\langle 0|C^2(\theta_{p_1})\cdots C^2(\theta_{p_m})C^3(\theta_{p_{m+1}})\cdots C^3(\theta_{p_{N}})|\Psi\rangle}
{\prod_{k=m+1}^N\Lambda(\theta_{p_k})}\no\\[6pt]
&&\hspace{-1.2truecm}\quad\quad =\sum_{1\leq p'_1<\cdots<p'_m\leq N}\frac{\langle \bar{0}|D^2_3(\theta_{p'_1})\cdots D^2_3(\theta_{p'_m})|\Psi\rangle}
{f_m(\theta_{p'_1},\cdots,\theta_{p'_m})\prod_{k=m+1}^N\Lambda(\theta_{p_k})}\no\\[6pt]
&&\hspace{-1.2truecm}\qquad\qquad \times
\langle 0|C^2(\theta_{p_1})\cdots C^2(\theta_{p_{m}})\,C^3(\theta_{p_{m+1}})\cdots C^3(\theta_{p_{N}})
\,D^3_2(\theta_{p'_1})\cdots D^3_2(\theta_{p'_m})|\bar{0}\rangle.
\eea
Substituting the equations (\ref{D-4}) and (\ref{D-5}) into the above equation,
we finally have the relation (\ref{Scalar-product-4}).

\end{document}